\documentclass[twocolumn,prl,showpacs,amsfonts,superscriptaddress,floatfix,nofootinbib]{revtex4-1}
\pdfoutput=1
\usepackage{amsmath}
\usepackage{graphicx}
\usepackage{footmisc}
\usepackage{color}

\graphicspath{ {fig2/} }

\begin{document}
\title {Topological states in the Hofstadter model on a honeycomb lattice}
\author{Igor N.Karnaukhov}
\affiliation{G.V. Kurdyumov Institute for Metal Physics, 36 Vernadsky Boulevard, 03142 Kiev, Ukraine}

\begin{abstract}
{
We provide a detailed analysis of a topological structure of a fermion spectrum in the Hofstadter model with different hopping integrals along the $x,y,z$-links ($t_x=t, t_y=t_z=1$), defined on a honeycomb lattice.
We have shown that the chiral gapless edge modes are described in the framework of the generalized Kitaev chain formalism, which makes it possible to calculate the Hall conductance of subbands for different filling and an arbitrary magnetic flux $\phi$. At half-filling the gap in the center of the fermion spectrum opens for $t>t_c=2^{\phi}$, a quantum phase transition in the 2D-topological insulator state is realized  at $t_c$. The phase state is characterized by zero energy Majorana states localized at the boundaries.
Taking into account the on-site Coulomb repulsion $U$ (where $U<<1$), the criterion for the stability of a topological insulator state is calculated at $t<<1$, $t \sim U$. Thus, in the case of $ U > 4\Delta $, the topological insulator state, which is determined by chiral gapless edge modes in the gap $\Delta$, is destroyed.
}
\end{abstract}
\maketitle

\section{Introduction}
In the framework of the Hofstadler model \cite{Hof}, we study the conductance of the 2D Chern insulator (CI), in which  the Hall conductance $\sigma_{xy}=\frac{e^2}{h}C_\gamma$ \cite{D1} is realized in the insulator state at the rational magnetic flux $\phi = \frac{p}{q}$, where p and q are relatively prime integers, $C_\gamma$ is the Chern number of $\gamma$-filled subbands. For  a square lattice and odd q, the spectral $q-1$ gaps are labeled by the Chern numbers.
The phase diagram representing the values of the Chern numbers as a function of a magnetic flux $\phi$ is known as the colored Hofstadter butterfly. The colored Hofstadter butterfly has been calculated for the honeycomb \cite{4}, triangular \cite{5} lattices. Wiegmann and Zabrodin  established that the Hofstadter model is related to the quantum group $U_q(sl_2)$ \cite{1},  the model Hamiltonian is determined in terms of the generators of the quantum group. Using the exact solution of the Hofstadter model defined on a square lattice for the rational flux \cite{1}, the authors \cite{2} have calculated the wave function of spinless fermions at zero energy in the semi-classical limit at $p=1$ and  $q \to \infty$. Near the boundaries of the sample the wave function of fermions is localized, power-low behavior of the modulus of the wave function is critical and unnormalizable. In the center of the spectrum, the chiral Majorana fermion liquid is realized at q=N (N is the size of the sample), the Bloch states disappear in the bulk of the system \cite {IK1}.  In \cite{a1,a2} the authors associate peculiarities of topological behavior at zero energy with the Van Hove anomalies, that exist in each band in the butterfly landscape.  In \cite{3} the solution \cite{1} has been generalized on the honeycomb lattice.

At half-filling the gapless state of the spinless fermions determined by the tight-binding model defined on the honeycomb lattice is unstable \cite{IK2}. In the case of taking into account the next-nearest-neighbor hoppings of fermions, nontrivial stable solutions for the phases, that determine these hopping integrals along the links, lead to spontaneous breaking of time reversal symmetry and CI state \cite{IK2}. We show, that in an external magnetic field and at half-filling the topological phase transition occurs from gapless state to gapped topological insulator state.

Topological insulators (superconductors) can be characterized alternatively in terms of bulk or edge properties, so-called the bulk-edge correspondence. So by calculating the winding number of the gapless edge modes along a loop in the Brillouin zone it is possible to assign to each gap its Chern number or the Hall conductance for given filling. In the case of an irrational flux the Brillouin zone is not defined, as a result, we can not calculate the Chern number integrating the Berry curvature over the Brillouin zone. We circumvent this problem by computing the Hall conductance using a different approach, calculating the total number of gapless edge modes in the gap \cite{A1}.  We show that the topological properties of the band spectrum reflect the nature of the corresponding chiral gapless edge modes localized at the boundaries of the sample, their total number in the gap is conserved under both rational and irrational magnetic fluxes.

In this paper we investigate the topological structure of the spectrum of spinless fermions defined on the honeycomb lattice. It is shown that the state of an insulator with nontrivial topological properties is realized in a magnetic field. This state is determined by zero-energy Majorana states localized at the boundaries of the 2D system. The proposed approach allows to calculate topological numbers in the case of irrational magnetic fluxes.
Despite numerous theoretical calculations, there is no universal criterion for realizing the topological state in the insulator with allowance for the short-range Coulomb repulsion. Such a criterion should connect the value of the gap of a topological insulator with the magnitude of the Coulomb repulsion between fermions.
Taking into account the exact solution of fermion chain model \cite{MN}, we calculate the stability of topological state in the Hofstadler model of interacting electrons. Result does not depend on the symmetry of the lattice, we believe that it is generic and can be applicable to different 2D topological insulators.

\section{Model Hamiltonian}

\begin{figure}[tp]
     \centering{\leavevmode}
    \begin{minipage}[h]{.8\linewidth}
\center{
\includegraphics[width=\linewidth]{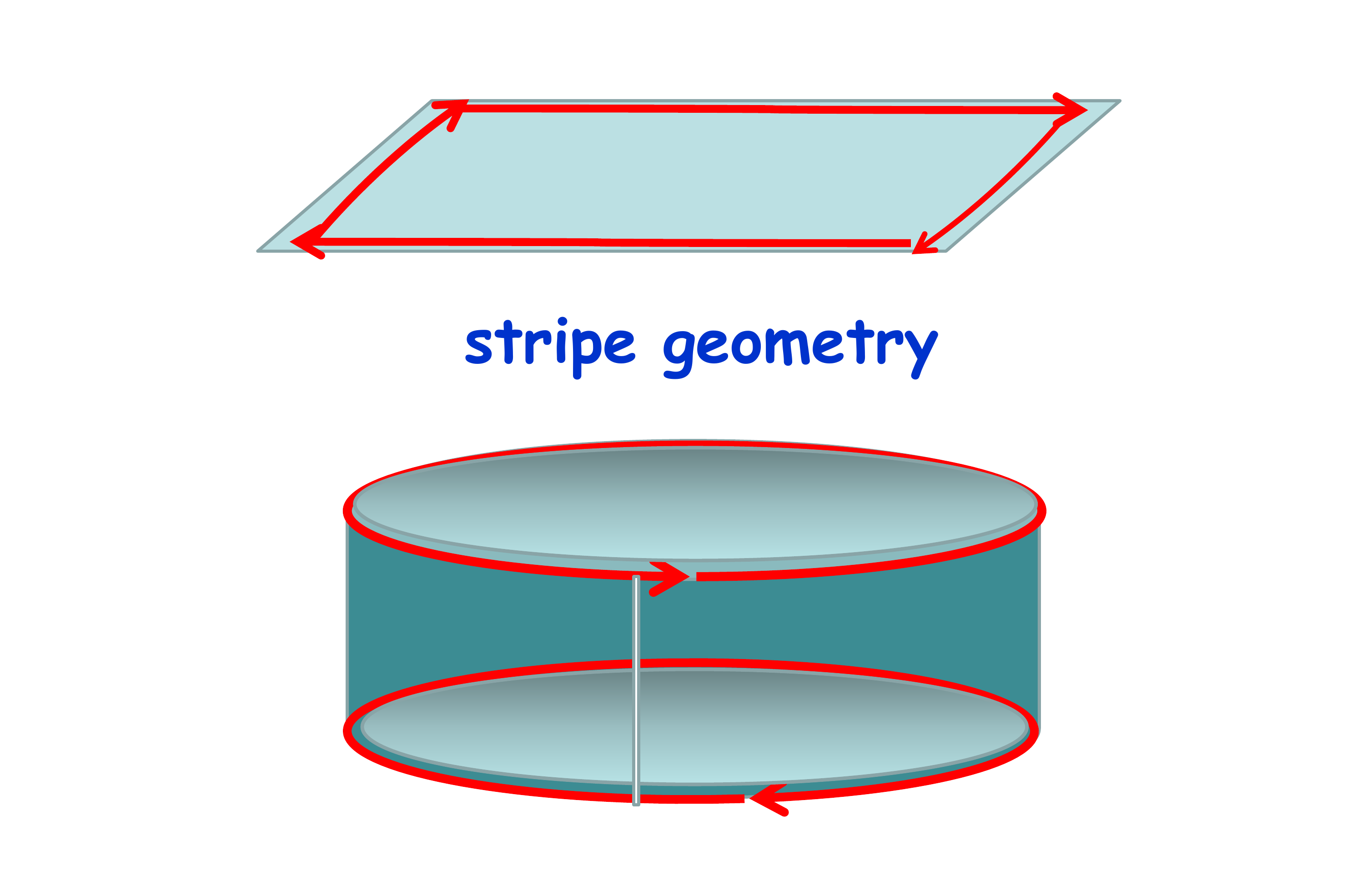}
                  }
    \end{minipage}
\caption{(Color online)
Stripe geometry for the 2D sample, chiral edge currents localized at the boundaries are marked with red arrows.}
\label{fig:0}
\end{figure}

We consider  CI insulator defined on a honeycomb lattice within the Hofstadter model \cite{Hof}. In the presence of a transverse homogeneous magnetic field $H{\textbf{e}}_z$ the model Hamiltonian has the following form
\begin{eqnarray}
&&{\cal H}_0= \sum_{x-links}\sum_{j} t^x(j) a^\dagger_{j}  a_{j+1}+ \sum_{y-links}\sum_{j} t^y(j) a^\dagger_{j} a_{j+1}+\nonumber \\
&& \sum_{z-links}\sum_{j} t^z(j) a^\dagger_{j} a_{j+1} + H.c.
    \label{eq:H0}
\end{eqnarray}
where $a^\dagger_{j} $ and $a_{j}$ are the fermion operators determined on a honeycomb lattice with sites $j$. The one-particle Hamiltonian (\ref{eq:H0}) describes the nearest-neighbor hoppings of spinless fermions with different amplitudes along the x-links $t^x (j)=t $ and along the y-and z-links  $t^y (j)=t^z(j)=\exp[i \pi (x_j-1)\phi]$. A magnetic flux through an unit cell $\phi = \frac{H }{ \Phi_0}$ is determined in the quantum flux unit ${\Phi_0=h/e}$, a homogeneous field~$H$ is represented by its vector potential $\textbf{A} = H x \textbf{e}_\xi$, $\textbf{e}_\xi=\{ \frac{1}{2}, \frac{\sqrt3}{2}\}$, while the X-axis coincides with the direction of the x-links, is perpendicular to the $\xi$-direction. The $x$-links determine coupling between zig-zag $\xi$-chains, the vector potential is directed along $\textbf{e}_\xi$. We consider the 2D fermion system in the stripe geometry Fig.\ref{fig:0} with open boundary conditions for the boundaries along the zig-zag chains (a sample has size N $\times$ N).  The magnetic field enters the Hamiltonian (\ref{eq:H0}) as the magnetic flux $\phi$ through an unit cell. The Hamiltonian is defined both rational and irrational magnetic fluxes.

\section{Topological structure of the spectrum}
\subsection{The rational flux $\phi =\frac{1}{4}$}
  \begin{figure}[tp]
    \centering{\leavevmode}
\begin{minipage}[h]{.9\linewidth}
\center{
\includegraphics[width=\linewidth]{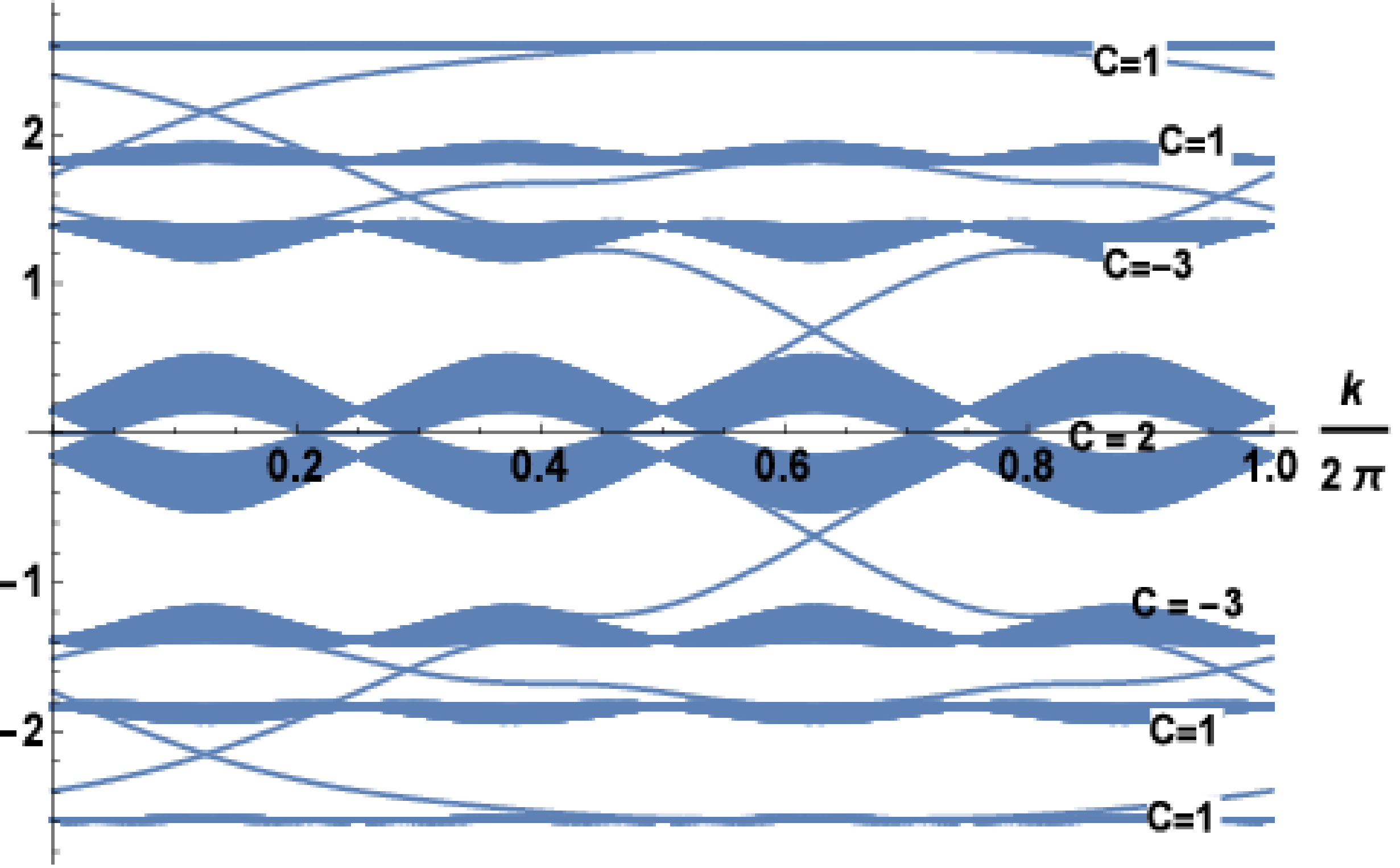} \\ a)
                  }
    \end{minipage}
       \centering{\leavevmode}
\begin{minipage}[h]{.9\linewidth}
\center{
\includegraphics[width=\linewidth]{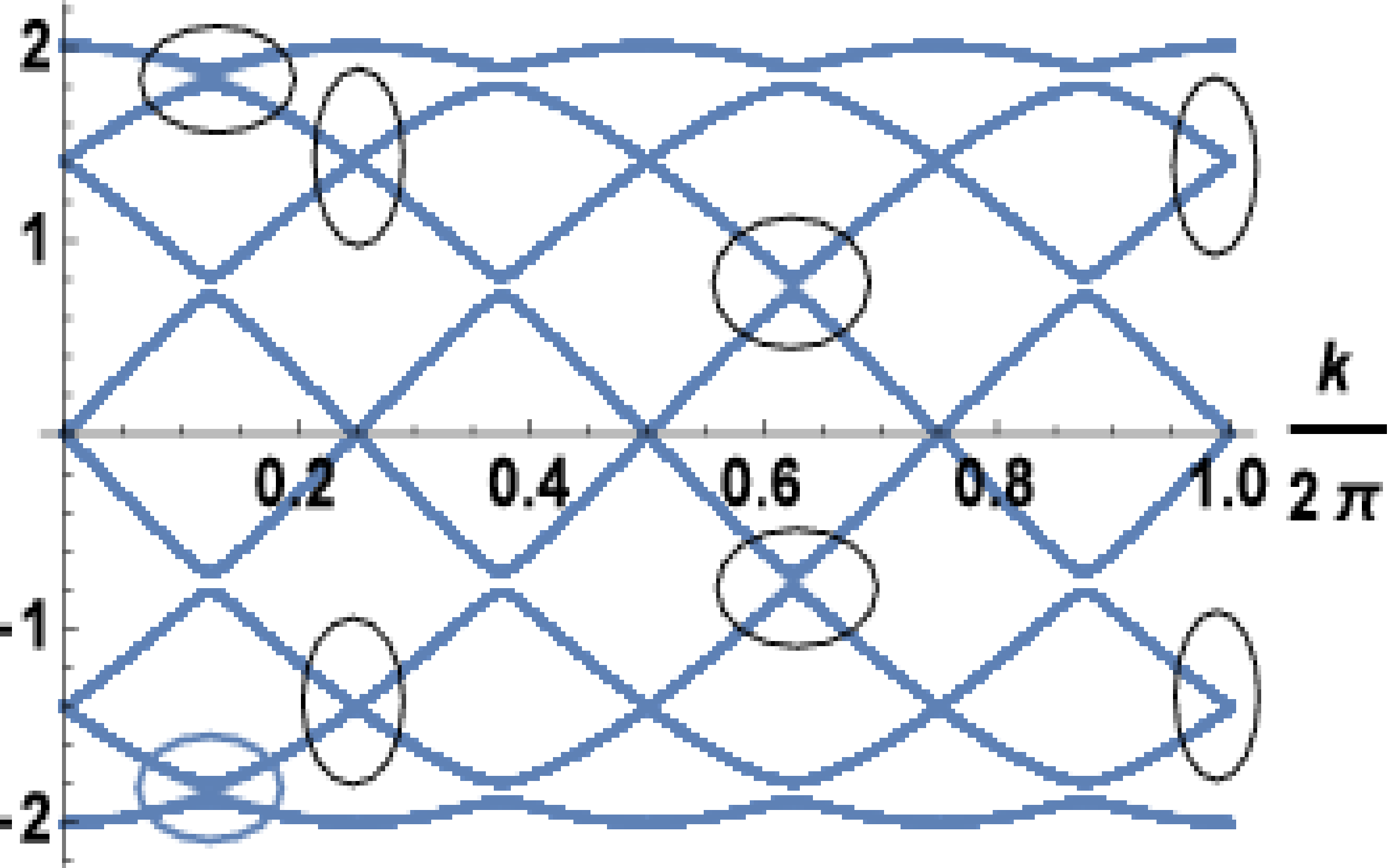} \\ b)
                  }
    \end{minipage}
     \centering{\leavevmode}
\begin{minipage}[h]{.9\linewidth}
\center{
\includegraphics[width=\linewidth]{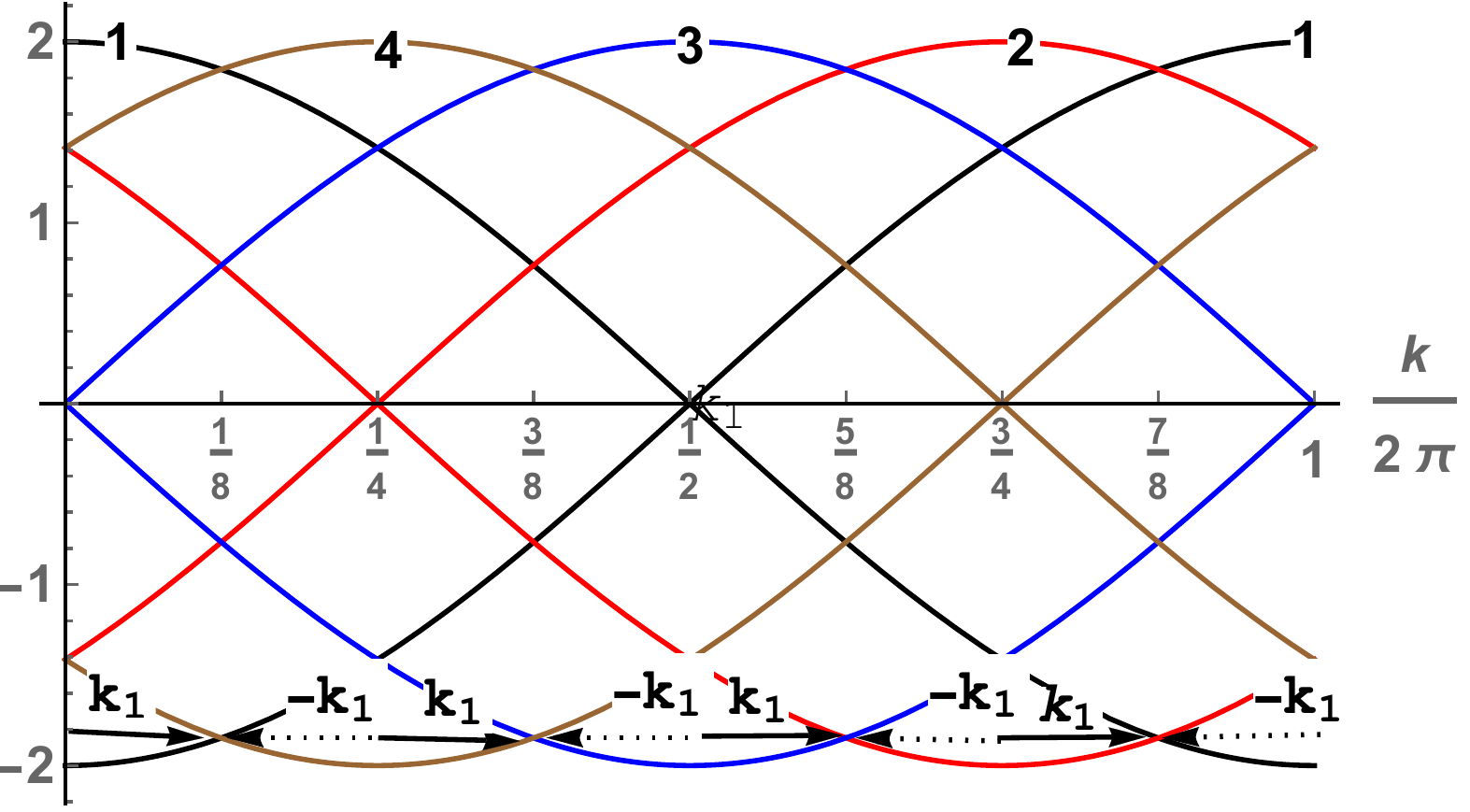} \\ c)
                  }
    \end{minipage}
\caption{(Color online)
Energy levels  calculated on a cylinder with open boundary conditions along zig-zag $\xi-$chains (the wave vector $k$ along the chain) for $\phi=\frac{1}{4}$, $t=1$ a), $t=\frac{1}{10}$ b) and $t=0$ c) ($k_1=\frac{\pi}{4}$). Colored lines indicate the numbered chains in the cell, ellipses show the gaps in which the chiral gapless edge modes are realized.
  }
\label{fig:1}
\end{figure}

\begin{figure}[tp]
     \centering{\leavevmode}
     \begin{minipage}[h]{.9\linewidth}
\center{\includegraphics[width=\linewidth]{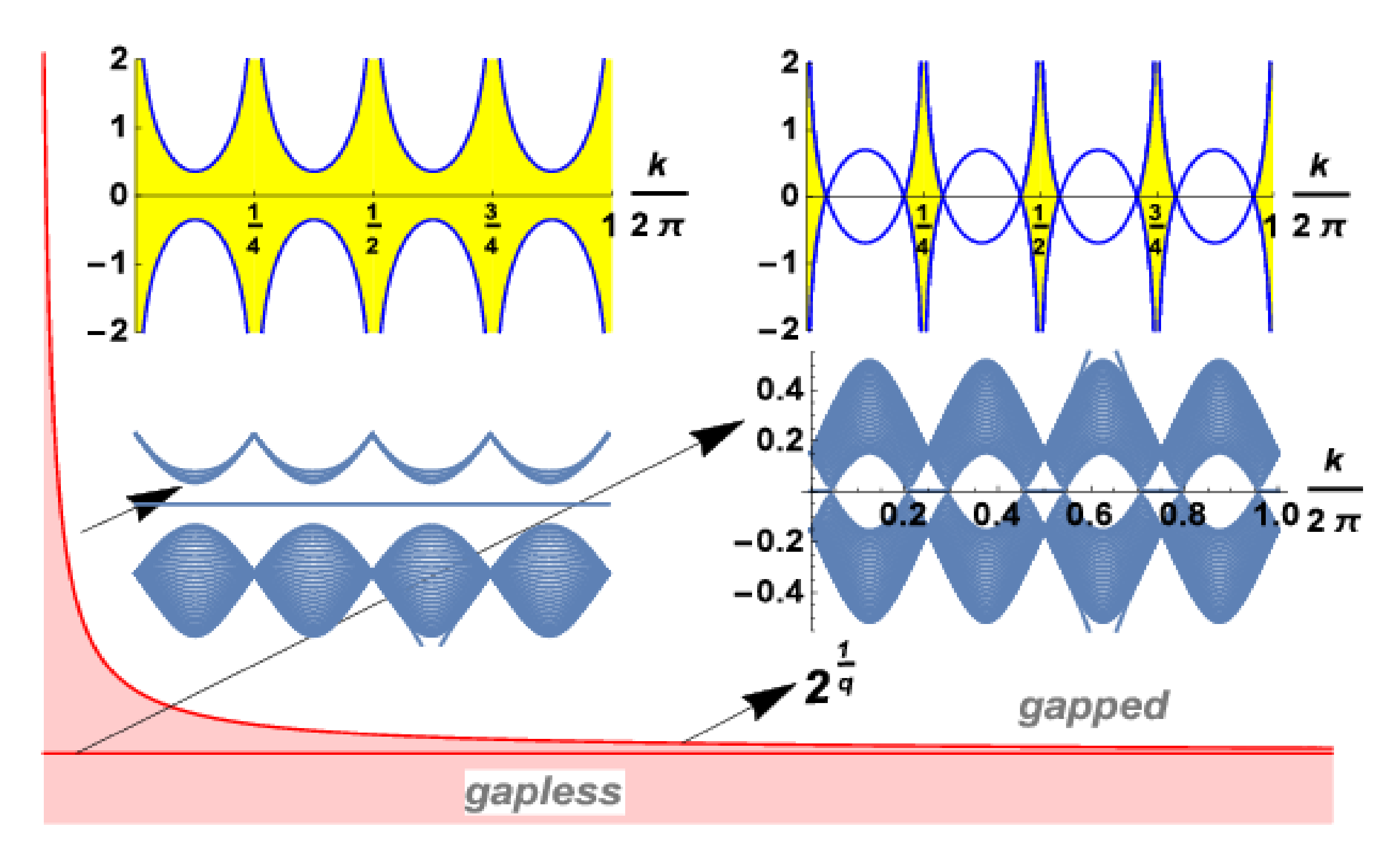}}
\end{minipage}
\caption{(Color online)
Phase diagram defined in the coordinates $q, t$ for half-filling (anisotropic hopping integral $t$, magnetic flux $\phi=\frac{1}{q}$),  $t=1$ corresponds to an isotropic case. The low energy spectra near the zero energy and the imaginary part of the wave vector $k_x$ as function of the wave vector along the $\xi$-chain, calculated at $q=4$, $t=1$ and $t=1.3$, illustrate the gapless and gapped states.}
\label{fig:2}
\end{figure}

Let us consider formation gaps in the butterfly landscape at a rational magnetic flux $\phi=\frac{1}{4}$. This case corresponds to a simple butterfly landscape in which the Chern number is different for different gaps. Numerical calculations of the fermion spectrum, obtained for a stripe geometry (see in Fig.\ref{fig:0}), are shown in Fig.\ref{fig:1}. The spectrum is symmetric with respect to zero energy, it includes seven isolated subbands. The topological structure of the subbands is determined by the following Chern numbers $\{1,1,-3,2,-3,1,1\}$, here two (gapless) subbands in the center of the spectrum are one topological subband with the Chern number $2$. Below we show that the state of fermions in the center of the spectrum (at $\epsilon =0$) is nontrivial for an arbitrary rational magnetic flux. At $\epsilon=0$ the spectrum is gapless for $t<t_c=2^{\frac{1}{4}}$, at $t>t_c$ the gap opens. We recall that for $\phi = 1 $ the gap opens at $t>2$ \cite{K2} (see in Fig.\ref{fig:2}), at weak next-nearest-neighbor hoppings of fermions the gap also opens \cite{IK2}. The gapless state is transformed into the gapped state at half-filling and $t>t_c$. From calculations of the spectra for an arbitrary rational flux $\phi =\frac{1}{q}$ it follows that $t_c=2^{\frac{1}{q}}$ (numerical calculations of the spectra for different $t$ were carried out in \cite{24}).

\begin{figure}[tp]
     \centering{\leavevmode}
     \begin{minipage}[h]{.95\linewidth}
\center{\includegraphics[width=\linewidth]{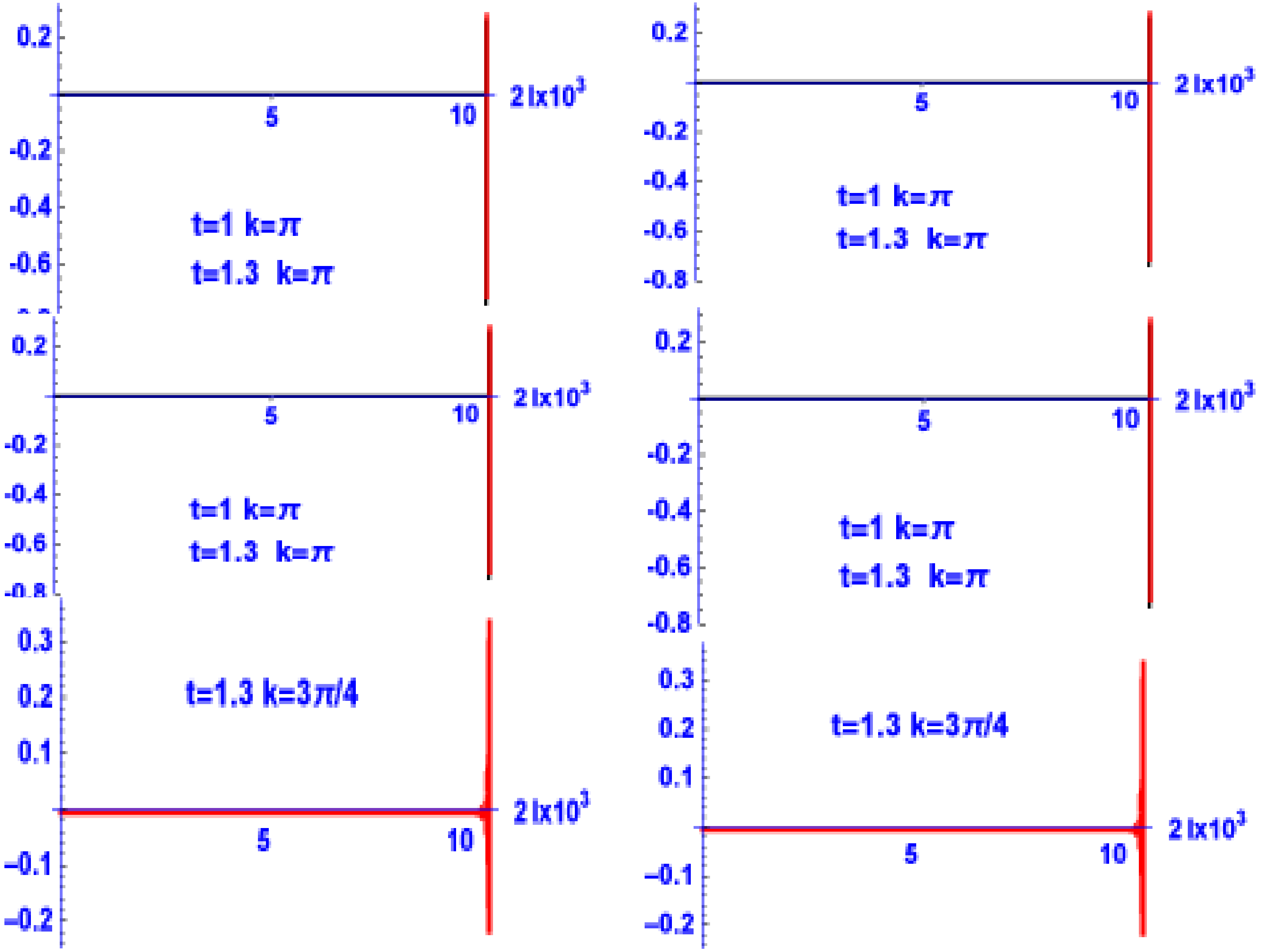}}
\end{minipage}
\caption{(Color online)
The real part of wave function of zero energy Majorana fermions (an imaginary part has the same behavior) defined on the lattice sites $l$ (along the $x$-direction, N=5000).  The edge modes are shown at $t=1$ for $k=\pi,0.92\pi$ and at $t=1.3$ for $k=\pi, 3\pi/4$, they are localized at the boundaries of the sample.
  }
\label{fig:3}
\end{figure}

Stability of zero energy states, localized at the boundaries, is determined by the imaginary part of the wave vector $k_x$ (wave vector with direction perpendicular to the boundaries), a change in its sign leads to the instability of this state. The imaginary part of $k_x$ depends on the value of the wave vector along the zig-zag chains $k$ and $t$, changes sign at the singular  points (see in Fig.\ref{fig:2}). At $t=1$ the points $p\frac{\pi}{2}\pm \frac{\pi}{12}$, $p=1,2,3,4$  define the regions of existence of zero energy Majorana states (marked by yellow) depending on the magnitude of the wave vector. These regions increase with $t$ and at $t>t_c$  zero-energy Majorana states  are realized in the entire range of $k$. For an arbitrary rational flux $\phi=\frac{1}{q}$ the bulk spectrum is $\frac{2\pi}{q} $-periodic. The wave function of zero energy Majorana states is localized at the boundaries of the sample, their behavior is the similar (but not the same) to chiral gapless edge modes that realized in CIs. At half-filling for $t<t_c$, the wave function is localized at different boundaries for the wave vectors $k$ which belong to the regions between the Dirac points. For the wave vectors outside these regions they are delocalized.  In the insulator state for $t>t_c$, the zero energy wave function is localized at different boundaries for any $k$ (see in Fig.\ref{fig:3}).

To calculate the edge states, we consider cylindrical system with bearded edges at both boundaries (the zig-zag boundaries along the $\xi$-direction). Zero energy Majorana states are realized at $\epsilon=0$, they are localized near the boundaries. The real part of the wave function of zero energy Majorana fermions as function of lattice site $l$ along the axis of cylinder is shown in Fig.\ref{fig:3}, the wave function is calculated for different $k$ (behavior of the imaginary part of the wave function is the same).
Thus, the 2D topological insulator state with zero energy Majorana fermion states is realized. At the same time, the Chern number and the Hall conductance are zero, which corresponds to trivial insulator state, in other words, the state of the 2D topological insulator with zero Chern number is realized.

Consider the behavior of the edge modes in the gaps \cite{IK}, which are realized at the energies $\epsilon=\pm 2 \cos \frac{\pi}{8}$  and $\epsilon=\pm 2 \sin \frac{\pi}{8}$ in the $t\to 0$ limit at $k_p$, here $k_p = \frac {(2p-1) \pi}{4}$ $p=1,...4$. Numerical calculations show that the values of the gaps $\Delta$ are equal to $\Delta=|t|+0(t^2)$. In the $t\to 0$ limit the energies of the zig-zag-$\xi$ chains crosse in the points $k_p$, the energies of the chains are shifted in $k_0= \frac{2\pi}{q}=\frac{\pi}{2}$. In Fig \ref{fig:1}c) it is shown, that two states of fermions of the $\xi-n$-chain with $k_n=k_0 (n-1)\pm k_1$ (where $k_1=\frac{\pi}{4}$), tunnel between the chains at the energy $\epsilon$. These tunneling processes are realized due to the conservation of the momentum and energy.
The corresponding states are determined by the Fermi operators $a^\dagger_n=a^\dagger (k_n),a_n=a(k_n)$. Taking into account the tunneling of fermions in the gaps  we can write the low-energy Hamiltonian ${\cal H}_{eff}= \tau(1) \sum_{n=1}^{N-1} a^\dagger_n a_{n+1}+ H.c.$, where $\tau (1)$ is a constant that determines the tunneling of fermions between the nearest-neighbor $\xi$-chains, $n$ define a lattice site along the $x$-direction or the number of the $\xi$-chain.  Using the hermitean and antihermitean parts of the $a_n$, leading to a Clifford algebra
$\chi_{n}=a_{n}+ a_{n}^\dagger$ and $\gamma_{n}=\frac{ a_{n}- a_{n}^\dagger}{i}$, we redefine the Hamiltonian as $i \frac{\tau (1)}{2} \sum_{n=1}^{N-1}( \chi_{n} \gamma_{n+1} -\gamma_{n} \chi_{n+1})$.  The Majorana operators $\gamma_n$ are defined by the algebra $\{\gamma_n,\gamma_m \}=2 \delta_{n,m}$ and $\gamma_n=\gamma^\dagger_n$.
Unlike traditional fermion hoppings, when the result of a hopping does not depend on the chirality (or the velocity of the particles), the conservation laws allow only hoppings of fermions between different chains with different velocities (for different signs of $k_1$ see in Fig.\ref{fig:1}c)) and with a given sequence of changes in the chirality of the movement of fermions along the $x$-chain (left-right, right-left,... for $\epsilon$ and right-left, left-right,... for $-\epsilon$).  We define the effective Hamiltonian, which takes into account the low energy excitations of Majorana fermions in the gap near the energy $\epsilon$
\begin{eqnarray}
{\cal H}(\epsilon) = i\frac{\tau(1)}{2}  \sum_{n=1}^{N-1} \chi_{n} \gamma_{n+1},
{\cal H}(-\epsilon) = i\frac{\tau(1)}{2} \sum_{n=1}^{N-1} \chi_{n+1} \gamma_{n}
\label{eq:H1}
\end{eqnarray}
In the case $N=q N_c$ ($N_c$ is the number of the q-unit cells) from the lattice periodicity $n \rightarrow n+q$,  we can consider the states of fermions of one q-cell or $N_c$ cells, taking into account the tunneling of fermions belonging the nearest-neighbor cells. The total Hamiltonian ${\cal H}(\epsilon)+{\cal H}(-\epsilon)$ does not break the $U(1)$ symmetry of the Hamiltonian (1).
According to Kitaev \cite{K1} the Hamiltonian (2) is describes the chain of isolated dimers of pairing Majorana fermions with the energy $\pm \frac{\tau (1)}{2}$ and two free zero energy Majorana fermions localized at the boundaries. The value of the gap, equal to $t$, corresponds to $\tau(1) \simeq t$.
At the given energies $\pm \epsilon$ two gapless edge modes with different chirality are localized at the boundaries, they determine the Hall conductance $\sigma_{xy}= (e^2 /h) C$ with the Chern number $C=1$. In the center of the spectrum at $\epsilon =0$, the Hamiltonians (\ref{eq:H1}) define gapless fermion states with the bandwidth $\tau(1)$ .

The generalization of the approach is based on the observation, that the Kitaev chains with the next-nearest neighbor hoppings between fermions describe two zero energy states of Majorana fermions localized at each boundaries. According to numerical calculations, the gaps are equal to $\Delta = t^2+0(t^3)$ at $\epsilon=\pm \sqrt2$. These gaps are formed due to the hoppings of fermions between the next-nearest neighbor $\xi$-chains (see in Fig.\ref{fig:1}). The structure of the effective Hamiltonian, which defines the states of fermions in the gaps is the same, it is determined as sum of two chains, defined on even and odd lattice sites
\begin{eqnarray}
&&{\cal H}(\epsilon) = \frac{i \tau(2)}{2} \sum_{n=1}^{N-2} \chi_n \gamma_{n+2}+ \frac{i \tau(2)}{2}\sum_{n=2}^{N-3} \chi_n \gamma_{n+2}, \nonumber \\
&&{\cal H}(-\epsilon) = \frac{i \tau (2)}{2} \sum_{n=1}^{N-2} \chi_{n+2} \gamma_{n}+ \frac{i \tau(2)}{2} \sum_{n=2}^{N-3} \chi_{n +2} \gamma_{n},
\label{eq:H2}
\end{eqnarray}
where tunneling constant $\tau (2)\simeq t^2$ determines the effective hopping between fermions located at the next-nearest neighbor $\xi$-chains.

The Hamiltonian (3) defines the zero energy states of Majorana fermions located at the ends of the chains with even and odd lattice sites (see in Fig.\ref{fig:1}). The values of the gaps in the spectrum depend on distance between $\xi-$chains $\delta$, the energies of which intersect for given $k$ and $\epsilon$. An effective tunneling constant for fermions of different chains is equal to $\tau(\delta)\simeq t^\delta+0(t^{\delta+1})$. The number of chiral edge modes localized at a boundary of the sample is equal to $\delta$, the dimension of a chiral mode is $\frac{1}{2}$ (they  propagate only in one direction). Calculating the energy of intersections of the spectra of the chains with the spectrum of the first chain and the distance between these chains and the first chain for an arbitrary rational flux $\phi=\frac{p}{q}$ at $t<<1$, we obtain the diophantine equation $p C_\gamma =q \cdot s +\gamma$ (here $s$ is an integer) \cite{4,24}, the same equation as for a square lattice \cite{D1,D2,20}. The Hall conductance $\sigma_{xy}=\frac{e^2}{h}C_\gamma$  is  defined for $\gamma$-filled subbands or fixed filling.

Finally, for given $\phi$ the equations for determining the Hall conductance in the gap $\Delta$, given with the energy $ \epsilon $, have the following form: $\epsilon(\delta)=\pm 2\cos(\pi\delta\phi/2)$ and $\epsilon(\delta )=\pm 2\sin(\pi\delta \phi/2)$ and $\epsilon(\delta )\neq 0$, these equations were obtained in the limit $ t \to 0 $, but they are valid for $t \leq 1$. The equations define the sequence of gaps $\Delta(\epsilon)$ in the fermion spectrum with their Hall conductance, when the Fermi energy lies in the gap $\Delta$. The spectrum is deformed with increasing $t$, but  the sequence of the gaps does not depend on the value of $t$. For a rational flux these equations reduce to the Diophantine equation presented above, where  $\gamma$ is the analog of $\epsilon$.

In the case of experimentally realizable magnetic fields, that corresponds to $q \sim 10^3-10^4, p=1$, in the semi-classical limit we obtain  $\epsilon (\delta)=\pi \frac{\delta}{q}$, $\delta=1,2,3,...$ for the states near the center of the spectrum. This result, obtained in the $t\to 0$ limit, valid for $t<t_c$. In the case of $t>t_c$ (see Fig.\ref{fig:2}), the lowest energy gap corresponds to a insulator state with $\sigma_{xy}=0$ at half filling. This set of the gaps corresponds to the levels with the energies $\epsilon (\delta)=\frac{\pi}{q} (\delta+1/2)$ \cite{19}. The experiments \cite{ex1,ex2} confirm an interesting behavior of the Hall conductance in graphene, which is quantized. This is an interesting new phenomena completely explained by the relativistic Dirac spectrum of graphene \cite{23}.

Let us try to explain the behavior of the Hall conductance in the semi-classical approximation. The regions of the spectrum near the edges of a band, that correspond to small filling of particles or holes, and the center of spectrum, that correspond to a half-filling, are separated. The fermion states that determined by the Landau levels with the energies $E=\pm[3-\sqrt{3}\pi\phi(n+1/2)]$ and the state with energies $\pm 3^{3/4}\pi \sqrt{2 \phi n}$ \cite{25} near the Dirac points have different the Hall conductance. At small filling the Chern number of n-filled subbands of spinless fermions is equal to n. Taking into account the spin degenerate and the Zeeman splitting we obtain the expression for the Hall conductance $\sigma_{xy}=\frac{e^2}{h}(2n+1)$.

\begin{figure}[tp]
     \centering{\leavevmode}
    \begin{minipage}[h]{.9\linewidth}
\center{\includegraphics[width=\linewidth]{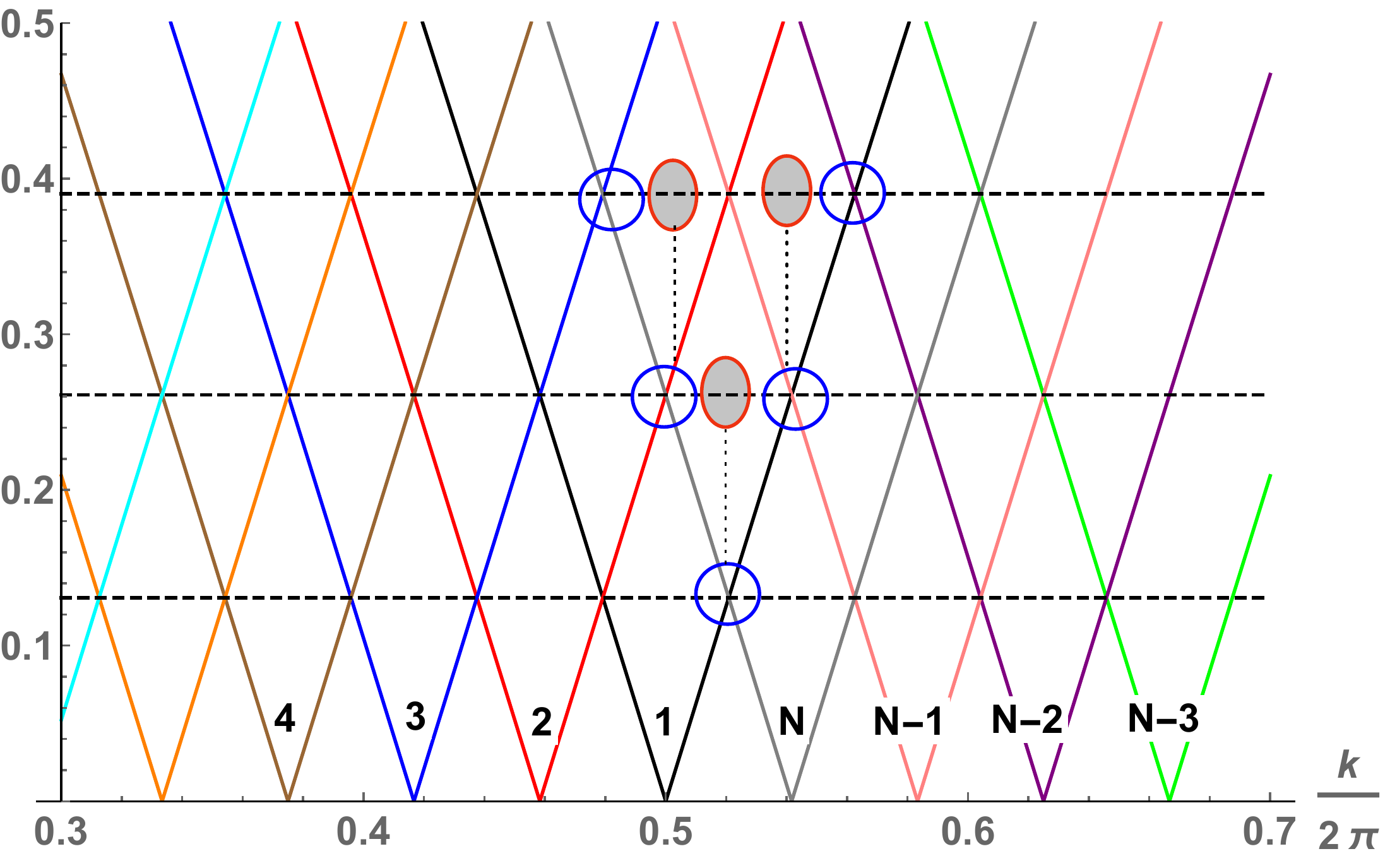}}
\end{minipage}
\caption{(Color online)
Energies of fermions (shifted by $\frac{2\pi}{q}$) in atomic layers at the boundaries (1;N) near the center of the spectrum as a function of the wave vector, the spectrum is symmetric around zero energy. Dashed lines define the Fermi energies or the center of the gaps in insulator states in $t\to 0$-limit in the case n=0,1,2. The chiral modes and the pairs of spinless fermions localized at the boundaries are denoted by circles and disks.
  }
\label{fig:4}
\end{figure}

For the fermion states near the center of spectrum the Zeeman splitting is small, so you can disregard it.
We must take into account the degeneracy of states with energies near the center of the spectrum. The lowest Dirac level, separated by a space, in which two gapless edges are localized at the boundaries at x = 1 and x = N, resulting in $ C_1 = 1 $. The Chern number for the next level determined by 'new' two chiral modes localized at x=2 and x=N-1 and modes localized at x=1 and x=N with filling 1 (they form the pair of spinless fermions with different chiralities and momenta near the point $k=\pi$ see in  Fig.\ref {fig:4}). For the Fermi energy, that corresponds the second gap, the states of gapless edge modes with  $k_- < k <k_+ $ ($k_\pm =\pi-\frac{2\pi}{q}\pm \frac{2\pi}{q}$) are fulfilled,
this state corresponds to a pair of spinless fermions. In this case the Chern number is equal to $C_2=1+2=3$, the structure of the Chern number $C_3$ is as follows $C_3= 1 +2 \times 2$. It is considered that two 'new' modes localized at x=3 and x=N-2 plus two pairs of spinless fermions at x=1,x=N and x=2,x=N-1. For the states of spinless fermions near the center of the spectrum we obtain the expression for the Chern number $C_n= 2n-1$ (here (n-1) is the number of pairs of spinless fermions localized at the boundaries), the Hall conductance is equal to  $\sigma_{xy}=\frac{e^2}{h}(2n-1)$ and $\sigma_{xy}=2\frac{e^2}{h}(2n-1)$ for electrons.

\begin{figure}[tp]
     \centering{\leavevmode}
    \begin{minipage}[h]{.9\linewidth}
\center{\includegraphics[width=\linewidth]{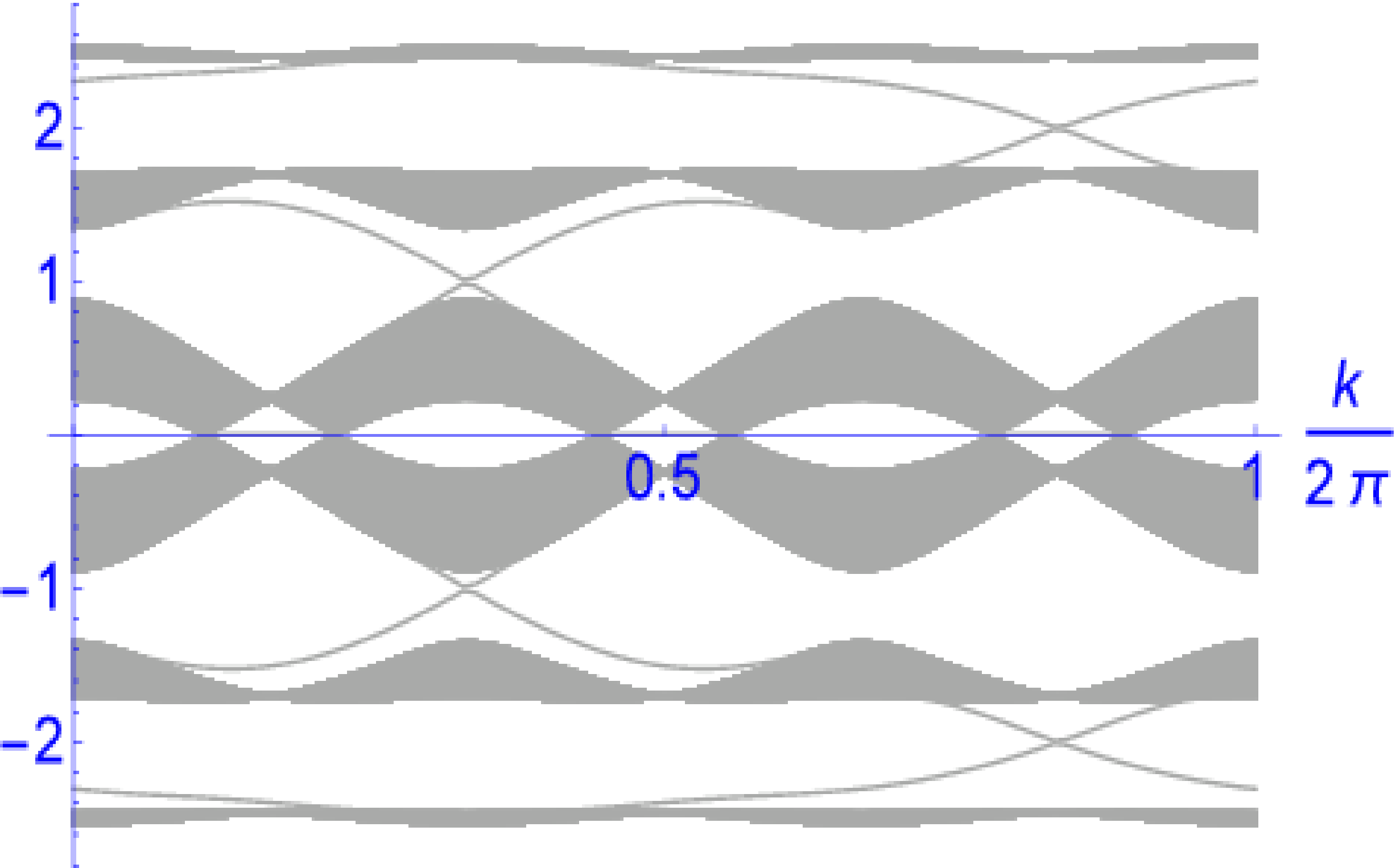} a)}
\end{minipage}
  \begin{minipage}[h]{\linewidth}
\center{\includegraphics[width=.9\linewidth]{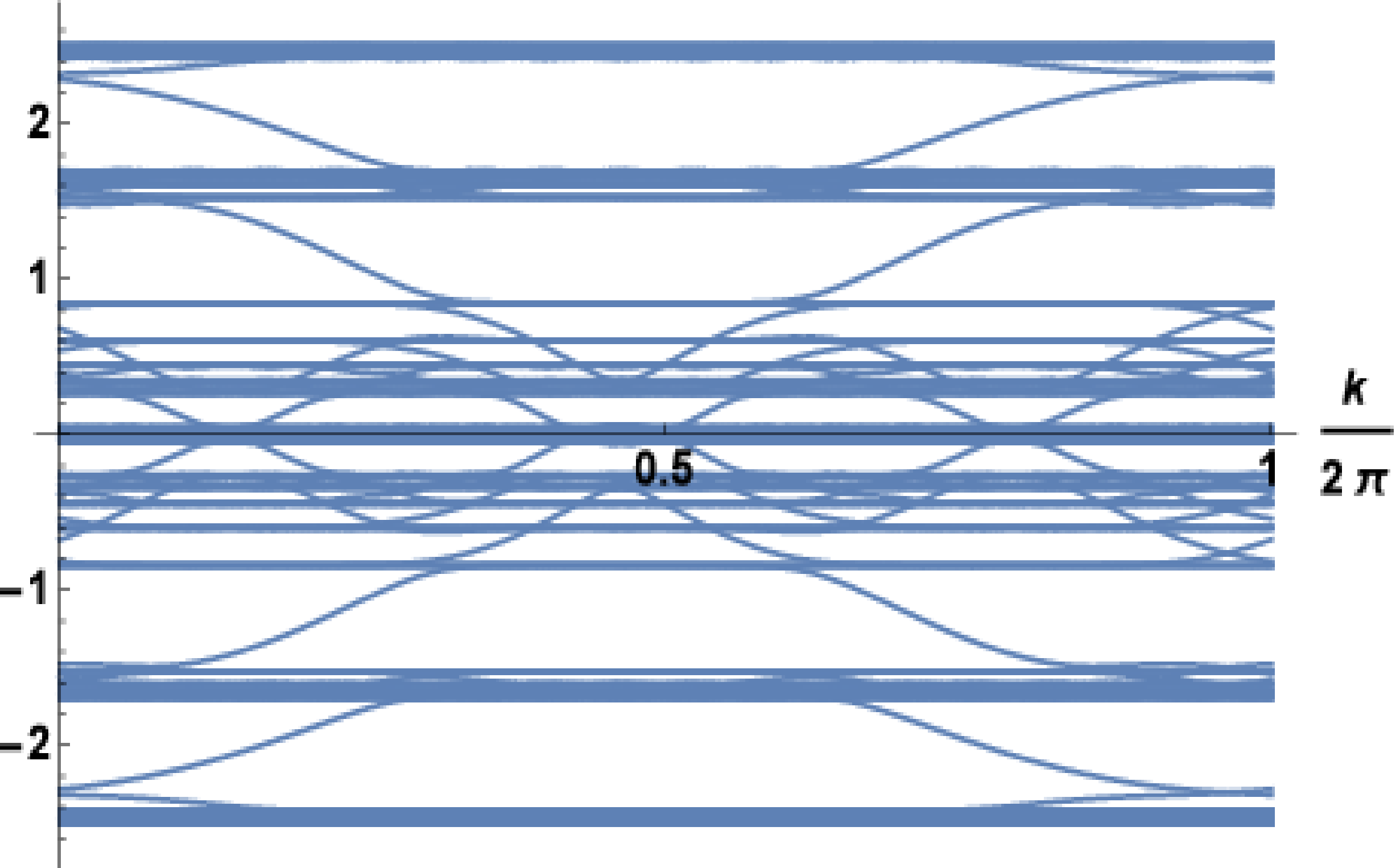} b)}
\end{minipage}
\begin{minipage}[h]{\linewidth}
\center{\includegraphics[width=.9\linewidth]{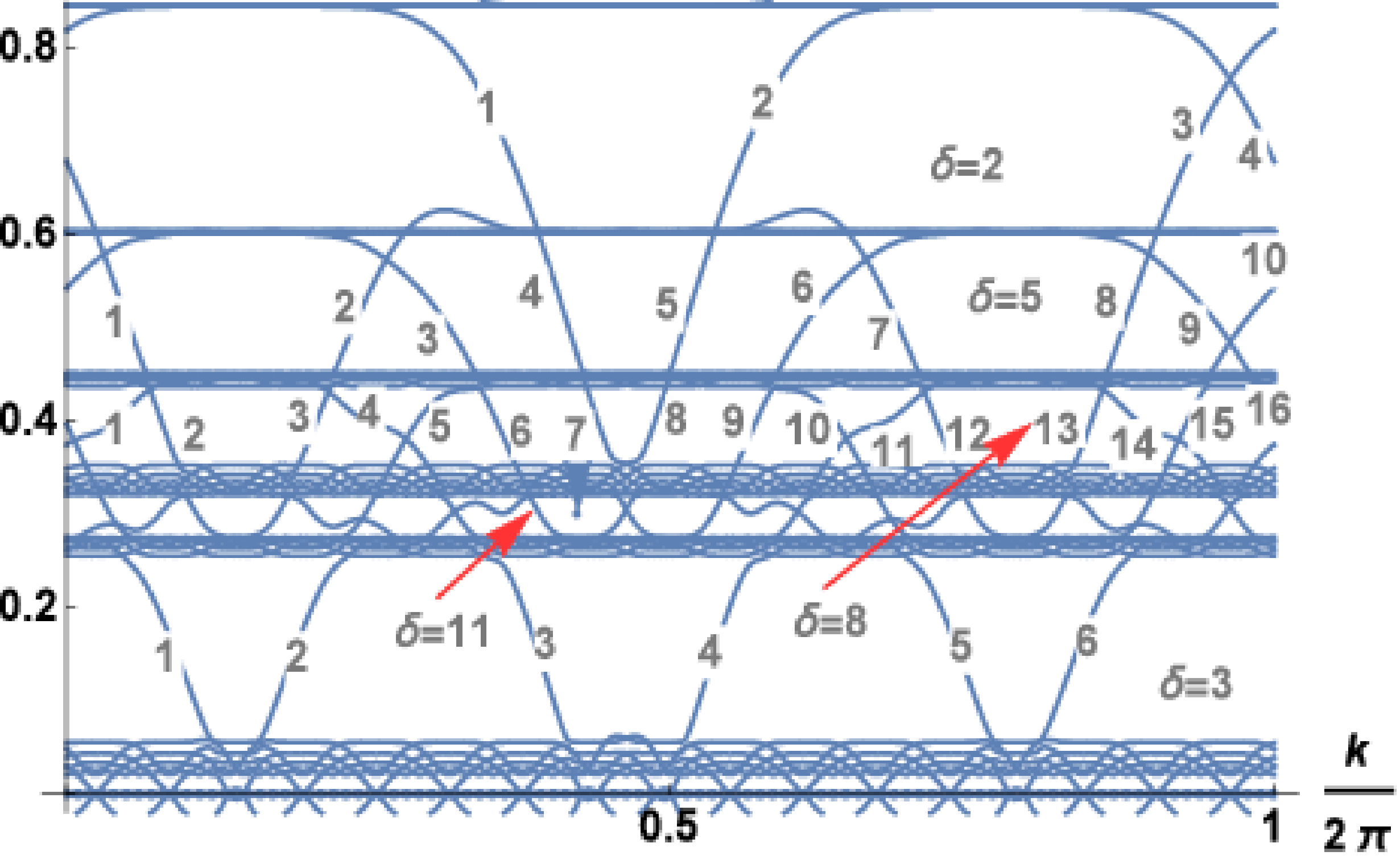} c)}
\end{minipage}
\caption{(Color online)
Energy levels  calculated on a cylinder with open boundary conditions along $\xi$-direction for $t=1$, $\phi=\frac{1}{3}$ a), $\phi=\frac{1}{\sqrt 8}$ N=200 b). The fine structure of the low-energy subband  c), $\delta$ denotes the total number of gapless edge modes localized at a boundary of the sample (the modes in the gaps are numerated).
  }
\label{fig:5}
\end{figure}

\subsection{The irrational flux $\phi=\frac{1}{\sqrt 8}$}

We study the spectrum and topological numbers in the case of an irrational magnetic flux. As an illustration, we consider the evolution of the simplest rational flux $q=3$ (see in Fig.\ref{fig:5}a)) to an irrational flux $q=\sqrt 8 =2.82843...$ (see in Fig.\ref{fig:5}b),c)). Although the flux values differ slightly, the fine structure of the spectrum with other topological numbers corresponds to the irrational flux.
For an arbitrary flux $\phi$ in the $t \to 0$ limit, the energies of fermions in the chains are shifted by $\pi \phi$ and intersect at the points $k_p=\pi p \phi$, $\epsilon_p=\pm 2 \cos (\frac{\pi p \phi}{2})$, $\epsilon_p=\pm 2 \sin (\frac{\pi p \phi}{2})$, here $p=1,...,\infty$. The set $k_p$, $\epsilon_p \neq 0$  determines the gaps in the fermion spectrum for given irrational flux $\phi$.

Numerical calculations of the excitation spectrum for a rational flux in the Hofstadter strips with open boundary conditions were obtained for samples of sizes $\frac{N}{q}$ is a natural number. When calculating the spectrum for an irrational flux, which is approximated by rational numbers $\phi \simeq \frac{p}{q}$, we assume that $N < q$. In this case numerical calculations correspond to a spectrum with an irrational flux to within $\frac{1}{N}$.
Calculations of the fermion spectrum for $\phi=\frac{1}{\sqrt 8}$, $t=1$, obtained on the lattice with size $N=200$, are shown in Fig.\ref{fig:5}b),c). Comparing the spectra with $q=3$ Fig.\ref{fig:5}a) and $q=\sqrt 8$ Fig.\ref{fig:5}b) we see its transformation. So a fine structure of each subband is formed with an irrational flux, the spectrum remains symmetric with respect to zero energy. A complex fine structure of the spectrum is realized near its center. We consider the fine structure of the subband near zero energy, limited in energy $1<\epsilon\leq 0$ (see in Fig.\ref{fig:5}c)). In the $t\to 0$ limit we obtain the following set of the gaps for the fine structure of the low energy subband $ \epsilon= 0.888 \Rightarrow \delta= 2$, $\epsilon = 0.714 \Rightarrow \delta=5$, $\epsilon =0.564 \Rightarrow \delta =9$, $\epsilon =0.533 \Rightarrow \delta =8$, $\epsilon= 0.379 \Rightarrow \delta=6$, $\epsilon = 0.347 \Rightarrow \delta=11$, $\epsilon=0.190 \Rightarrow \delta=3$,..., where the energy $\epsilon$ determine the position of the gap in the fine structure of the subband and $\delta$ is the number of gapless edge modes localized at a boundary of the sample. We have considered all possible gaps in the fine structure of the low energy subband when $ \delta $ changes from 1 to 11.
The topological properties of the fine structure of the subband are conserved with an increase in the value of $t$, therefore it is fair to expect that the Hall conductance of each subband is determined by these values of $\delta$.
The calculation of the fine structure of a low energy subband is shown in  Fig.\ref {fig:5} c), the energies corresponding to the gaps vary, but their sequence and the total numbers of gapless edge modes in the gaps, that define the Hall conductance of the subbands, are conserved. The fine structure of the subband is determined by the gaps with $\delta =2, 5, 8, 11, 3$ (see in Fig.\ref {fig:5} c)). A super fine structure determines the splitting of subbands with very close energies $\epsilon =0.564$ and $\epsilon =0.533$, $\epsilon= 0.379$ and $\epsilon= 0.347$, it is the band collapse in the case of an irrational flux \cite{19}.

\section{The Hofstadter model of interacting electrons}

The interaction between fermions 'kills' the state of the topological isolator, it is obvious that in the case of strong interaction a trivial insulator state is realized. Below, we consider in detail the stability of the topological insulator state  in the presence of the on-site interaction $U\sum_{j}n_{j;\uparrow} n_{j;\downarrow}$. In graphene the distance between the Landau levels is very large compared to the Zeeman splitting \cite{23}, so we will not consider this term.  The total Hamiltonian takes into account also the terms (\ref{eq:H0}) for fermions with different spins. In the limit $t \to 0$ the model (1) with the on-site Hubbard interaction is reduced to the isolated Hubbard $\xi-$chains. We consider the case of a weak on-site interaction $U << 1 $, when $U \sim t$ or $U \sim \Delta$. In the case of a weak interaction the electron spectrum of the Hubbard $\xi$-chain is renormalized slightly.

We accent our consideration on the behavior of edge modes in the gaps for an rational flux. As we noted above, only two states of spinless fermions into the gaps with fixed wave vectors $\pm \pi\delta/q$, defined at one lattice site $n$, tunnel between the $\xi-$chains, these states have different chirality (see in Fig.\ref{fig:1}c)). Taking into account that the density of fermions at the site $j=\{n,l\}$ ($l$ numerates the lattice sites along the $\xi$-chains) and the energy $\epsilon$ is determined as $n_{n,l;\sigma}=n_{n;\sigma}(\epsilon)=\frac{i}{2}\gamma_{n;\sigma} \chi_{n;\sigma}+\frac{1}{2}$.
Taking into account that $\sum_{l}( n_{n,l;\uparrow}-\frac{1}{2})(n_{n,j;\downarrow}-\frac{1}{2})=\frac{1}{N}\sum_{k} m (n,k;\uparrow) m(n,-k;\downarrow)$,  where $m(n,k;\sigma)=\sum_{l}(n_{n,l;\sigma}-\frac{1}{2})\exp (i k l)$, we obtain $\frac{1}{N}\sum_{k} m(n,2k_0(n-1);\uparrow)m(n,2k_0(n-1);\downarrow)=
- \frac{1}{4}\gamma_{n;\uparrow}\chi_{n:\uparrow} \gamma_{n;\downarrow}\chi_{n:\downarrow}$,
here $m(n,2 k_0(n-1);\sigma)=n_{n;\sigma}(\epsilon)-\frac{1}{2}$. We used that the spectrum is periodic with the period $k_0$. Taking into account the interaction term, we define an effective low-energy Hamiltonian \begin{equation}
{\cal H}_{eff} = i\frac{\tau (\delta)}{2} \sum_{\sigma} \sum_{n=1}^{N-\delta}\gamma_{n;\sigma} \chi_{n+\delta;\sigma}
-\frac{U}{4}\sum_{n}  \gamma_{n;\uparrow} \chi_{n;\uparrow}\gamma_{n;\downarrow} \chi_{n;\downarrow},
\label{eq:H4}
\end{equation}
where $\delta$ denotes also the number of the chains of electrons with the hopping integral $\tau(\delta)$ between electrons located at the sites on the distance $\delta$, in (4) Majorana fermions are defined for fermions with different spins.

Taking into account the spin freedom of fermions, the Chern number for electron subbands and the gapless edge modes  doubles.  The Hamiltonian (4) with the on-site Hubbard interaction has be diagonalized exactly by  Mattis and Nam \cite{MN}. The model is reduced to chain of spinless fermions (2) with the chemical potential $U$  \cite{MN}.  According to \cite{MN}, the ground state degeneracy is dependent on whether $\kappa=\frac{U}{\tau(\delta)}>4$ or otherwise. In the thermodynamic limit there are zero energy Majorana states, they are realized in the interval $|\kappa|\leq 4$. The number of chiral gapless edge modes is associated with the topological order, so for $|\kappa |=4$  in the thermodynamic limit a topological phase transition between phases with different Chern numbers is realized. The value $\tau (\delta)$ decreases with increasing of $\delta$, depends on $\epsilon$.  The phase state, for which $ |\frac{U}{\Delta}|>4$, is topological trivial state, when the interaction is taking into account. The topological ambitions of the subbands are limited by a weak interaction $U<4\Delta$.

\section{Conclusion }

We have studied the behavior of 2D fermions in the Hofstadter model, defined on the honeycomb lattice, focusing our attention on the structure of the chiral gapless edge modes in the gaps. In contrast to the traditional approach, when the states of the bulk fermions are mainly taken into account, we pay attention to the study of fermion states in the insulator gaps, since these states determine the Hall conductance. This approach makes it possible to calculate the Hall conductance in CI for an arbitrary flux. The structure of chiral gapless edge modes is described within the framework of the Kitaev chain, with effective hopping integral between Majorana fermions.
For an irrational flux the fine structure of the subbands transforms to the super fine structure with infinite number of subbands and different numbers of chiral gapless edge modes.
We have shown also, that at half-filling a gapless state, is unstable in the case an isotropic Hofstadter model; an external magnetic field leads to gapped (insulator) state.
The 2D topological insulators that support chiral gapless edge modes are extremely susceptible to short range electron-electron interactions, for $ U > 4 \Delta$, the Hubbard interaction destroys the Majorana gapless edge states. We hope, that the obtained criterion for the stability of the topological state with allowance for the short-range repulsion between fermions is applicable to a wide class of topological insulators.

This research was partially supported by the budget program 6541230 "Support for the development of priority areas of research"


\begin{thebibliography}{31}
\bibitem{Hof} D.Hofstadter, {Energy levels and wave functions of Bloch electrons in rational and irrational magnetic fields}, { Phys.Rev.B}, 14 (1976), 2239-2249. https://doi.org/10.1103/PhysRevB.14.2239
\bibitem{D1} I.Dana, Y.Avron and J.Zak, {Quantised Hall conductance in a perfect crystal},  {J.Phys.C: Solid State Phys.}, 18 (1985), L679-684. https://doi.org/10.1088/0022-3719/18/22/004
\bibitem{4} A.Agazzi, J.-P.Eckmann and G.M.Grafm,
{The colored Hofstadter butterfly for the honeycomb lattice}, {J.Stat.Phys.}, 156 (2014), 417-426.  https://doi.org/10.1007/s10955-014-0992-0
\bibitem{5} J.E.Avron,  O.Kenneth and G,Yehoshua,
{A study of the ambiguity in the solutions to the Diophantine equation for Chern numbers},
{J.Phys.A: Math.Theor.}, 47  (2014), 185202. https://doi.org/10.1088/1751-8113/47/18/185202
\bibitem{1} P.B.Wiegmann and A.V.Zabrodin,  {Bethe-ansatz for the Bloch electron in magnetic field}, {Phys.Rev.Lett.} 72, (1994), 1890-1893.  https://doi.org/10.1103/PhysRevLett.72.1890
\bibitem{2} Y.Hatsugai, M.Kohmoto and Y.-S. Wu,  {Explicit solutions of the Bethe ansatz equations for Bloch electrons in a magmnetic field}, {Phys.Rev.Lett. }, 73 (1994), 1134-1137. https://doi.org/10.1103/PhysRevLett.73.1134
\bibitem{IK1} I.N.Karnaukhov, {Chern insulator with large Chern numbers. Chiral Majorana fermion liquid}, \emph{J.Phys.Commun.}, 1 (2017), 051001.   https://doi.org/10.1088/2399-6528/aa9541
\bibitem{a1} G.G.Naumis, {Topological map of the Hofstadter butterfly: Fine structure of Chern numbers and Van Hove singularities}, Phys. Lett  A. 380 (2016), 1772. https://doi.org/10.1016/j.physleta.2016.03.022
\bibitem{a2} I.I.Satija, Topology and self-similarity of the Hofstadter butterfly, arXiv:1408.1006 [cond-mat.dis-nn] (2014).
\bibitem{3} M.Eliashvili, G.I.Japaridze  and G.Tsitsishvili,  { The quantum group, Harper equation and the structure of Bloch eigenstates on a honeycomb lattice}, {J.Phys.A:Math.Theor.}, 45 (2012), 395305. https://doi.org/10.1088/1751-8113/45/39/395305
\bibitem{IK2} I.N.Karnaukhov, {Spontaneous breaking of time-reversal symmetry in topological insulators}, {Physics Letters A}, 381 (2017), 1967-1970. https://doi.org/10.1016/j.physleta.2017.04.014
\bibitem{A1}
J.C.Avila, H.Schulz-Baldes and C.Villegas-Blas,{Topological invariants of edge states for
periodic two-dimensional models}, {Math. Phys. Anal. Geom. }, 16 (2013), 137-170.  https://doi.org/10.1007/s11040-012-9123-9
 \bibitem{MN} D.C.Mattis and S.B.Nam,  {Exactly soluble model of interacting electrons}. {J.Math.Phys.}, 13 (1972), 1185-1189. https://doi.org/10.1063/1.1666120
\bibitem{K2}A.Yu.Kitaev, {Anyons in an exactly solved model and beyond}, {Annals of Physics}, 321 (2006), 2-111. https://doi.org/10.1016/j.aop.2005.10.005
\bibitem{24} M.Sato, D.Tobe and M.Kohmoto, {Hall conductance, topological quantum phase transition, and the Diophantine equation on the honeycomb lattice}, {Phys.Rev.B}, 78 (2008), 235322. https://doi.org/10.1103/PhysRevB.78.235322
\bibitem{K1} A.Yu.Kitaev, {Unpaired Majorana fermions in quantum wires}, {Phys.- Usp.}, 44 (2001), 131-136. https://doi.org/10.1070/1063-7869/44/10S/S29
\bibitem{IK} I.N.Karnaukhov, {Edge modes in the Hofstadter model of interacting electrons},
    \emph{Europhysics Letters}, 124 (2018), 37002. https://doi.org/10.1209/0295-5075/124/37002,
    see also C.Malciu, L.Mazza, and C.Mora, \emph{$4\pi$ and $8\pi$ dual Josephson effects induced by symmetry defects}, arXiv:1901.03342v1 [cond-mat.mes-hall] 2019.
\bibitem{D2} I.Dana, {Topologically universal spectral hierarchies of quasiperiodic systems}, {Phys.Rev.B} \textbf{89} (2014) 205111. https://doi.org/10.1103/PhysRevB.89.205111
\bibitem{20} G.Amit and I.Dana, {Topological phase transitions from Harper to Fibonacci crystals}, {Phys.Rev.B}, 97 (2018), 075137. https://doi.org/10.1103/PhysRevB.97.075137
    \bibitem{19} B.A.Bernevig, T.L.Hughes, S.-C.Zhang, H.-D.Chen and C.Wu, {Band collapse and the quantum Hall effect in graphene}, {Int.J.Mod.Phys.B}, 20 (2006), 3257-3278. https://doi.org/10.1142/S0217979206035448
\bibitem{ex1}
K.S.Novoselov, A.K.Geim,  S.V.Morozov,  D.Jiang, M.I.Katsnelson, I.V.Grigorieva, S.V.Dubonos and A.A.Firsov,  {Two-dimensional gas of massless Dirac fermions in graphene}, { Nature}, 438 (2005), 197-200. https://doi.org/10.1038/nature04233
\bibitem{ex2}
Y.Zhang, Y.W.Tan, H.L.Stormer  and P.Y.Kim, {Experimental observation of the quantum Hall effect and Berry's phase in graphene}, { Nature}, 438 (2005), 201-204. https://doi.org/10.1038/nature04235
\bibitem{23} V.P.Gusynin and S.G.Sharapov, {Unconventional integer quantum Hall effect in graphene},{ Phys.Rev.Lett.} 95 (2005), 146801. https://doi.org/10.1103/PhysRevLett.95.146801
\bibitem{25}
Y.Hatsuda, {Perturbative/nonperturbative aspects of Bloch electrons in a honeycomb lattice}, arXiv:1712.04012v2 [hep-th] (2018).
\end{thebibliography}
\end{document}